\documentstyle[twoside,fleqn,espcrc2]{article}

\newcommand{\AmS}{{\protect\the\textfont2
   A\kern-.1667em\lower.5ex\hbox{M}\kern-.125emS}}

\title{The glueball among the light scalar mesons}

\author{Peter Minkowski\address{Institute for Theoretical Physics,
     University of Bern,  CH-3012 Bern, Switzerland}
         \thanks{Work supported in part by Schweizerischer Nationalfonds}
         and
         Wolfgang Ochs\address{Max Planck Institut f\"ur Physik, Werner
      Heisenberg Institut,
            D-80805 Munich, Germany}}

\begin{document}

\begin{abstract}
 In our phenomenological analysis
of the spectroscopy of light scalar mesons we do not find
compelling evidence for the existence of the low mass $\kappa(900)$ or
$\sigma(600)$ states nor for
$f_0(1370)$ as single resonance. If the $f_0(980)$ and
and $f_0(1500)$ are taken as members of the $q\overline q$ nonet there
remains a broad object formed by $f_0(400-1200)$ and $f_0(1370)$ 
which is a glueball candidate $gb(1000)$.
\end{abstract}

% typeset front matter (including abstract)
\maketitle

\section{Introduction}
The existence of glueballs is among the fundamental predictions of QCD,
but the experimental evidence is still in
debate. In QCD calculations on the lattice in quenched approximation 
the lightest glueball is found to have quantum
numbers $J^{PC}=0^{++}$ and a mass in the region 1400-1800 MeV 
(for review, see \cite{bali}). The effect of this approximation is 
still being investigated.
In an alternative approach based on QCD sum
rules a gluonic state of lower mass around 1000 MeV is required as
well \cite{narison}. 

The search for the lightest glueball should therefore
concentrate on the mass region up to about 1800 MeV in the scalar
sector. This search has to proceed in parallel with the identification
of the low mass scalar $q\bar q$ nonet(s). The glueball candidate should
fulfill some general properties, it should be produced in particular in a
gluon rich environment and its decay (for the unmixed glueball) should be
``flavour-blind''. Then the scalar states to be identified
as members of the $q\bar q$ nonet or glueball should be found from the list
provided by  the Particle Data Group\cite{pdg}\\
I=0: $f_0$(400-1200), ($\sigma(600)$?), $f_0(980)$, $f_0(1370)$,    
\hspace*{8mm}  $f_0(1500)$, $f_0(1710)$\ldots\\
I=$\frac{1}{2}$: ($\kappa(900)$?), $K^*_0(1430)$, $K^*(1950)$\ldots\\
I=1:  $a_0(980)$, $a_0(1450)$\ldots\\

There are different scenarios for interpretation which include
Scenario A:\\
A starting point is the lattice result, then a glueball with suitable mass
is $f_0(1500)$ \cite{ac}. As this state does not obey all wanted  properties
one includes the nearby $f_0(1370)$ and $f_0(1710)$ and all three scalars
mix with
the glueball and two $q\bar q$ states (for review of this popular
approach, see \cite{klempt}). In this case the low mass states like
$a_0(980)$ and $f_0(980)$, possibly also $\kappa$ and $\sigma$, are
considered as multiquark states which should not be included in the 
$q\bar q$ spectroscopy.\\
Scenario B:\\
One tries to identify the $q\bar q$ nonet first including the $a_0$ and
$f_0$. In our approach \cite{mo,mo2} the $f_0(980)$ and
$f_0(1500)$ are the isoscalars of the $q\bar q$ nonet with large mixing,
very similar to the $\eta,\ \eta'$ in the pseudoscalar sector. The nonet is
completed with $a_0(980)$ and $K^*_0(1430)$. The $\sigma$ and $\kappa$
are not considered as genuine resonances. The remaining states called
$f_0(400-1200)$ and $f_0(1370)$ (``red dragon'' )
correspond to a single broad state which we suggest to be
the lightest glueball. 
The same scalar states are
chosen on the basis of a theoretical model in \cite{instanton}, except for
the $a_0(980)$, 
the existence of a glueball is not accepted though \cite{klempt}.  
A similar scheme with a broad state as glueball is considered in
\cite{anisns},
albeit at a higher mass; in this scheme the $f_0(1370)$ is taken as $q\bar
q$. A glueball around 1000 MeV which mixes with a nonstrange isoscalar
 is also obtained in the
QCD sum rule approach \cite{narison}.

In order to distinguish these possibilities it is of primary importance to
verify the existence of the states in question and to determine their
constituent structure.
 As relevant criterion for
a resonant state to be included in spectroscopy we require the amplitude to
describe a full circle in the complex plane (``Argand diagram''), possibly
distorted by a smoothly varying background amplitude. This requires the
verification of the appropriate phase motion of the amplitude through
an analysis of angular distributions.

\section{Is there a scalar $K\pi$ resonance $\kappa(900)$?}
Important new information to this discussion has been provided by the
FOCUS collaboration on the semileptonic decay $D^+\to \mu^+\nu K^-\pi^+$
\cite{focus}. From the forward backward asymmetry in the $K\pi$ decay 
they conclude on the presence of an S-wave component
in the mass region studied from 0.8 to 1.0 GeV with constant
phase $\phi_S=45^\circ$ interfering with a
$K^*$ Breit-Wigner resonance. Because of the Watson theorem this phase
should equal the elastic $K\pi$ scattering phase. This 
has been determined through
the study of $K^-p\to K^-\pi^+n$ by isolating the one-pion-exchange
contribution \cite{lass}. Indeed, the two results closely agree, 
the elastic phase shift
slowly rises from $\sim 35^\circ$ to $\sim 50^\circ$ in the mass range
considered. An S-wave resonance $\kappa(900)$ with width around
400 MeV 
would lead to a phase variation of $\sim 50^\circ$  in this mass region 
apparently not observed by FOCUS.
The elastic $K\pi$ amplitude in \cite{lass}
has been parametrized by a superposition of the $K^*_0(1430)$ and a 
background as $S=T(K^*_0)e^{2i\delta_B} +T_B$ where the background phase
 rises slowly and reaches $\delta_B\sim 50^\circ$ at the $K^*_0$. This
background we do not consider as evidence for an additional resonance  
to be included in spectroscopy (see also \cite{cp}).
 
A related process is the hadronic decay $D^+\to K^-\pi^+\pi^-$ studied by
the E791 Collaboration\cite{E791}. The Dalitz plot shows again the presence
of the $K^*(890)$ and $K^*_0(1430)$. If the S wave is fitted with 
 an energy independent background amplitude and phase factor multiplying
$K^*$ no satisfactory description is obtained.
With an additional $\kappa$ resonance at mass 798 MeV
and width 410 MeV a better fit is obtained. Such a result appears to
contradict the above FOCUS result with almost constant phase in the region
800-1000 MeV. We expect that a better fit can be obtained as well if the
elastic $K\pi$  phase shifts are used, parametrized by a
background phase and phase factor for the $K^*_0$ both energy dependent 
as in \cite{lass}. 

As an additional check of the fit to the Dalitz plot we suggest comparing
not only to the mass spectra with fine binning but also to the higher
moments of the decay angular distribution, in particular the first moment
$\langle Y_1^0\rangle\sim \langle {\rm cos} \theta\rangle$ where cos$\theta$,
 for given $m^2(K^-\pi^+_1)$,
 is related to  $m^2(K^-\pi^+_2)$ (see also \cite{wo}). These moments 
should reflect the angular
distribution of the $K^-\pi^+_1$ channel but with a smooth backround
from the resonances in the $K^-\pi^+_2$ channel. Inspection of the Dalitz
plot suggests again a strong variation of the asymmetry
 $ \langle {\rm cos} \theta\rangle$ over the $K^*(890)$ region. 
For the time being we see no compelling evidence for
an additional $K\pi$ resonance below 1 GeV.
   
\section{Is there a scalar  $\pi\pi$ resonance $\sigma(600)$ ?}
The elastic $\pi\pi$ phase shifts are by now rather well known and a 
unique solution is established up to $\sim $ 1400 MeV \cite{klr},
similar to the old results \cite{cm}.
%These experimental phase shifts are also found consistent
%with requirements from analyticity and unitarity as realized in Roy
%equations \cite{acgl}. These phase shifts show the rapid phase variation
%due to $f_0(980)$. 
If the rapid phase variation due to  $f_0(980)$ is removed
one finds a slowly moving
phase passing $90^\circ$ at around 1000 MeV and another 
more narrow structure presumably related to $f_0(1500)$. 
The behaviour of this phase
shift may be parametrized by a Breit-Wigner resonance of 1 GeV mass 
with large width of at least 500 MeV \cite{mo}; if additional background is
included the resonance position may shift to higher values around 1400 MeV
\cite{anisns} or lower values around 600 MeV \cite{iii}. The question may be
asked whether in other reactions, with different background,
a resonance around 600 MeV appears with a corresponding phase variation.
Several such proposals have been put forward and we consider some of them 
in the ~following.

1. $J/\psi\to \omega\pi\pi$\\
There is a peak around 500 MeV in the $\pi\pi$ mass spectra
which may be a signal from a  $\sigma$ Breit-Wigner resonance
\cite{dm2}.
Then the interference term Re$(SD^*)$
between the (almost real) D wave which is dominated 
by $f_2(1270)$ and the resonant S wave would change sign
at the mass of the $\sigma$ and so
the angular distribution
$d\sigma/d\Omega
\sim |S|^2 + 10 (3\cos^2\vartheta-1)$ Re$(SD^*) + {\cal O}(|D|^2)$
would vary
accordingly with a sign change of the $\cos^2\vartheta$ term (from + to --).
The data \cite{dm2}
do not show any sign change below 750 MeV and therefore there is
no indication for a Breit Wigner resonance at 500 MeV.

2.  Central $\pi\pi$ production in $pp\to p(\pi\pi)p $\\
The mass distribution of the pion pair in this double Pomeron dominated
process peaks shortly above threshold $\sim 400$ MeV \cite{afs} and has been
related to the $\sigma(600)$ as well. However, again, there is no related
phase variation of the S wave amplitude which should become visible from the
$S-P$ or $S-D$ interference terms. In this process we understand the origin
of the peak. We propose this process to be dominated at low masses by
Pomeron Pomeron $\to \pi\pi$ through one pion exchange very much like
$\gamma\gamma\to \pi \pi$, the latter process is discussed in \cite{bp}. 
Indeed there is a close similarity between these two processes: the $I=0$
component peaks below 400 MeV and the D-wave already near 500 MeV 
with 1/3 of intensity. 
As the pion pole is near the physical
region the  $\pi \pi$ angular distribution is very steep, steeper than
in more typical
interactions mediated by vector ($\rho$) exchange. Therefore
one estimates that the D wave
becomes important not at $m_{f_2}$ but already
at $m_{f_2} \times (m_\pi/m_\rho) \sim 0.3$ GeV. This mechanism also
explains the low mass peak of the S wave without associated phase variation.
The production of a broad state at 1000 MeV as in elastic $\pi\pi$
scattering is possible in addition either by rescattering or by direct
formation as in $\gamma\gamma \to \pi\pi$ \cite{bp}.

3. Decay $D^+\to \pi^-\pi^+\pi^+$\\
The $\pi^+\pi^-$ mass spectrum presented by the
E791 Collaboration \cite{E791a} shows three prominent peaks,
one just above $\pi\pi$ threshold, one related to  $\rho$ and one to
 $f_0(980)$.  Only
fits including a light $\sigma$ particle have been found successful
according to their analysis. In analogy to the decay $D^+\to K^-\pi^+\pi^+$
discussed above we would expect the low mass region to be governed by the
elastic $\pi\pi$ scattering phase without additional resonance
contributions. This should apply strictly to the corresponding semileptonic
decay $D^+\to \pi^-\pi^+\mu^+\nu$. As discussed above for the $\kappa$
a resonant $\sigma(600)$ should yield a characteristic interference pattern
in the projected $\langle {\rm cos}\theta\rangle$ moment which should be analysed. For the
moment we consider the question of the $\sigma(600)$ in this process as
open.  

In conclusion, the first two processes show peaks at low mass but definitely
no resonant phase motion, for the latter process this question is not
definitely answered. 
We do not discuss here other peaks related to $\sigma\to \pi\pi$
($\psi'\to J/\psi \pi \pi$,  $Y',Y''\to Y\pi\pi$, $\tau\to \nu_\tau 3\pi$,
$f_0(1370/1500)\to 4\pi$) which are all lacking a phase analysis. 
For the moment we suppose  the $\pi\pi$ phase shifts in all these processes 
could behave as in elastic $\pi\pi$ scattering 
with a broad state around 1 GeV but
no narrower state below 1 GeV. 

\section{Has $f_0(1500)$ a strong glueball component?}
The main argument \cite{mo} against a strong glueball 
component in  $f_0(1500)$ is the observed negative relative phase
between the amplitudes
\begin{equation}
T(\pi\pi\to  f_0 \to K\overline K)= -
T(\pi\pi\to f_0 \to \eta\eta).
\label{phase1500}
\end{equation}
These amplitudes have been reconstructed from the measured $|S|,|D|$,
their relative phase and the absolute phase of the D wave resonances.
A similar behaviour of amplitudes 
(although with different overall phase) 
is also found in the fits by \cite{anisph}. This constraint strongly
restricts the possible admixture of a glueball component which 
would contribute
with the same sign to all pseudoscalar particle pairs. In particular,
among the six models listed in Table 4 of \cite{klempt} 
which describe the $f_0(1500)$ as superposition of
$u\bar u+d\bar d,\ s\bar s$ and glueball only two \cite{lw,cel}
yield a negative sign between the amplitudes (\ref{phase1500}) with
a glueball amplitude of 0.22 and 0.01 resp. or a probability of less than
4\% for the glueball component. 

\section{The candidate scalar glueball}
Having constructed the low mass $q\bar q$ nonet with $f_0(980),f_0(1500),
a_0(980)$ and $K^*(1430)$ \cite{mo,mo2} 
and not accepting $\sigma$ and $\kappa$ 
as genuine resonances below 1 GeV the states left in in our list at low mass
are $f_0(400-1200)$ and $f_0(1370)$. We do not discuss here $f_0(1370)$,
a full circle in the Argand diagram has not been established in our view
\cite{mo,wo}, also there are problems with an inconsistency of branching
ratios \cite{klempt}.
These two objects we consider as
representing one single broad resonance as suggested in particular by
elastic $\pi\pi$ scattering and this is our glueball candidate $gb(1000)$.

This state fulfils most standard requirements on glueballs:\\
1. central  production in $pp$ scattering;\\
2. decay of radially excited states $\psi',Y',Y''$ into the respective
ground state and $\pi\pi$, this can only proceed through gluonic
intermediate states;\\
3. Low energy $p\bar p\to 3\pi$;\\
4. $J/\psi\to \gamma \pi\pi$: no prominent signal is observed here up to now
  which is the only unfavourable aspect of our hypothesis;\\
5. Flavour properties: The decays of the glueball component $f_0(1370)$
 favours a glueball over a nonstrange $q\bar q$ assignment (see, for
example, \cite{seth});\\
6. Small coupling to $\gamma\gamma$, see our discussion \cite{mo1};
7. t-channel analysis of elastic $\pi\pi$ scattering:
% Whereas 
%the $I_t=1$ amplitude has little background, the $I_t=0$ amplitude 
%has large background already below 1 GeV and 
The isoscalar exchange amplitude below 1 GeV cannot be saturated
by $q\bar q$ resonances alone \cite{mo1}.

\section{Conclusions}
We found no definitive evidence for the low mass $\kappa$ and $\sigma$
Breit-Wigner resonances below 1 GeV, in particular, the semileptonic $D$
decays speak against the $\kappa$; similar clarity on the $\sigma$ 
could be obtained from semileptonic decays into $\pi\pi$. The $f_0(1500)$
can only have a small glueball component. The broad state $gb(1000)$ is
a realistic glueball candidate. Our analysis is mainly phenomenological
and driven by simplicity, this does not exclude more complex scenarios, such
as a mixing between $f_0(980)$ and $gb(1000)$. Further experimental results
would be helpful, for example the comparative study of quark and gluon
jets in their respective fragmentation regions \cite{mo3}.

\end{document}